\documentclass{emulateapj}

\newcommand{\myemail}{quanz@astro.phys.ethz.ch}

\slugcomment{To appear in ApJL}

\shorttitle{NACO/APP planet searches around HD172555 and HD115892 at 4-$\mu$m}
\shortauthors{Quanz et al.}

\begin{document}


\title{Searching for gas giant planets on Solar System scales: \\VLT NACO/APP observations of the debris disk host stars HD172555 and HD115892}

\author{Sascha P. Quanz}
\affil{Institute for Astronomy, ETH Zurich, Wolfgang-Pauli-Strasse 27, 8093 Zurich, Switzerland}    
\email{\myemail}

\author{Matthew A. Kenworthy}
\affil{Sterrewacht Leiden, P.O. Box 9513, Niels Bohrweg 2, 2300 RA Leiden, The Netherlands}  
\author{Michael R. Meyer}
\affil{Institute for Astronomy, ETH Zurich, Wolfgang-Pauli-Strasse 27, 8093 Zurich, Switzerland}    

\author{Julien H. V. Girard}
\affil{European Southern Observatory, Alonso de C\'ordova 3107, Vitacura, Cassilla 19001, Santiago, Chile}
\author{Markus Kasper}
\affil{European Southern Observatory, Karl Schwarzschild Strasse, 2, 85748 Garching bei M\"unchen, Germany.}        

\altaffiltext{1}{Based on observations collected at the European Organisation for Astronomical Research in the Southern Hemisphere, Chile, under program number 060.A-9800(J).}


\begin{abstract}
Using the APP coronagraph of VLT/NACO we searched for planetary mass companions around HD115892 and HD172555 in the thermal infrared at 4 $\mu$m. Both objects harbor unusually luminous debris disks for their age and it has been suggested that small dust grains were produced recently in transient events (e.g., a collision) in these systems. Such a collision of planetesimals or protoplanets could have been dynamically triggered by yet unseen companions. We did not detect any companions in our images but derived the following detection limits: For both objects we would have detected companions with apparent magnitudes between $\sim$13.2--14.1 mag at angular separations between 0.4--1.0$''$ at the 5-$\sigma$ level. For HD115892 we were sensitive to companions with 12.1 mag even at 0.3$''$. Using theoretical models these magnitudes are converted into mass limits. For HD115892 we would have detected objects with 10--15  M$_{\rm Jup}$ at angular separations between 0.4--1.0$''$ (7--18 AU). At 0.3$''$ ($\sim$5.5 AU) the detection limit was $\gtrsim$25 M$_{\rm Jup}$. For HD172555 we reached detection limits between 2--3 M$_{\rm Jup}$ at separations between 0.5--1.0$''$ (15--29 AU). At 0.4$''$ ($\sim$11 AU) the detection limit was $\gtrsim$4 M$_{\rm Jup}$. Despite the non-detections our data demonstrate the unprecedented contrast performance of NACO/APP in the thermal infrared at very small inner working angles and we show that our observations are mostly background limited at separations $\gtrsim$0.5$''$.

\end{abstract}



\objectname{HD172555}
\objectname{HD115892}


\section{Introduction}
While most dedicated surveys to directly image extrasolar planets around nearby stars yielded null results \citep[e.g.,][]{chauvin2010,heinze2010,lafreniere2007,kasper2007}, some remarkable exceptions were discovered in the last years: 
The HR8799 planetary system \citep{marois2008,marois2010}, Fomalhaut b \citep{kalas2008}, $\beta$ Pictoris b \citep{lagrange2009a,lagrange2010}, and 1RXS J1609-2105 b \citep{lafreniere2008,lafreniere2010}. The host stars of the first three systems are all A-type stars and they harbor both massive planets and debris disks. For Fomalhaut and $\beta$ Pictoris dynamical interactions between the exoplanets and the disks led to observable signatures: an offset between the debris disk center and the star (Fomalhaut) or disk warps ($\beta$ Pictoris). Thus, although there seems to be no direct correlation between the existence of both a debris disks and (an) exoplanet(s) \citep[e.g.,][]{apai2008,moro-martin2007}, specific properties of the debris disk can hint toward the existence of low-mass companions. 

Here, we report on the search for low-mass companions around HD115892 and HD172555 using direct imaging. Both objects are also young, nearby, A-type stars (see, Table~\ref{targets}) that are surrounded by debris disks \citep{oudmaijer1992,moor2006,su2006,morales2009}. These objects are, however, particularly interesting because their disks have a very high fractional luminosity and appear too luminous for their age \citep{moor2006,wyatt2007}. Since steady state evolutionary models of debris disks predict a much lower dust luminosity it was suggested that either the planetesimals in these systems have unusual properties or the observed, luminous dust could be transient \citep{wyatt2007}. Interestingly, \citet{lisse2009} found evidence for Silica dust and SiO gas in the mid-infrared spectrum of HD172555 which could be indicative of a high-velocity collision of protoplanets or planetesimals. Such a collision could have been triggered via dynamical interactions or gravitational stirring by a so far unseen companion, so we sought to search for it directly. 

\begin{figure}[!b]
\centering
\epsscale{.5}
\plotone{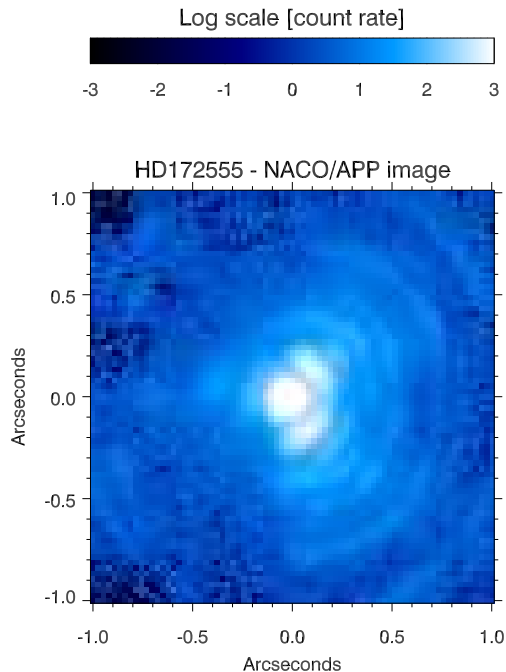}
\caption{The PSF of HD172555 observed with NACO/APP. The image shows the median combination of one cube of unsaturated exposures. In the left-hand side the APP effectively suppresses the diffraction rings between $\sim$2--7 $\lambda$/D.
\label{psf}}
\end{figure}

We used the Apodizing Phase Plate (APP) coronagraph installed at VLT/NACO  \citep{kenworthy2010,girard2010}. The APP is designed to work in the 3--5 $\mu$m wavelength range where it enhances the contrast between $\sim$2--7 $\lambda/D$ on one side of the PSF \citep[Figure~\ref{psf}; see also,][]{kenworthy2007,codona2006}. This inner working angle (IWA) corresponds to projected separations of $\sim$5--25 AU around our targets, comparable to the giant planets' orbits in our own Solar System. The debris disks are located within the inner 10 AU around each object (Table~\ref{targets}).


\section{Observations and data reduction}\label{observations_section}
The data were obtained on 2010-04-04 during the commissioning of the APP with the high-resolution AO-camera NACO 
\citep{lenzen2003,rousset2003} mounted on ESO's VLT UT4. Using the same observing setup already used to image the exoplanet $\beta$ Pictoris b \citep{quanz2010}, we chose the L27 camera ($\sim$ 27.15 mas pixel$^{-1}$) with the visible wavefront sensor. All images were taken in the NB4.05 filter ($\lambda_{c}=4.05\,\mu$m, $\Delta\lambda=0.02\,\mu$m) and in pupil stabilized mode. These are the first data sets that combine the APP with Angular Differential Imaging (ADI) \citep{marois2006} (see below). We used the "cube mode" readout where all image frames, i.e., each single exposure, are saved individually. To ensure that no frames were lost we only read out a 512$\times$512 pixels sub-array of the detector. The effective field-of-view (FoV) using the APP in this sub-array mode is restricted to the uppermost 512$\times$90 pixels (i.e., roughly 13.9$''\times$2.4$''$) as the APP introduces a vertical shift of the image along the detector's y-axis\footnote{See, NACO User Manual v. 88 page 53.}. For both sources, several data cubes were taken each at a slightly different dither position following a  3--point dither pattern along the x-direction of the detector's effective FoV. Halfway through the observations the camera was rotated by 180$^\circ$ so that the high contrast side of the APP covered both hemispheres. Due to an error during the rotation of the camera, we did not cover the full 360$^\circ$ around the targets (see, section~\ref{analysis}).
In total we obtained 54 and 36 data cubes for HD115892 in hemisphere 1 and 2, respectively, and 18 and 24 data cubes for HD172555. Each cube consists of 200 individual image frames, i.e., exposures. Table~\ref{observations} summarizes the observations and also the observing conditions. To enhance the signal-to-noise (S/N) of potential companions, we chose to saturate the core of the stellar PSFs, but we note that the APP reduces the peak flux in the PSF core by roughly 40\% \citep{kenworthy2010}.

For the photometric calibration we also obtained unsaturated images of both targets. We used the same observing strategy but decreased the detector integration time (DIT) to 0.0558 s for HD115892 and 0.2 s for HD172555 and took only 6 data cubes, each consisting of 200 exposures, for each calibration data set. In Figure~\ref{psf} we show the median combined image of 1 cube for HD172555. 


The general data reduction approach (bad pixel correction, sky subtraction) is described in \citet{quanz2010}. This time, however, since we had sufficient field rotation ($>$10$^\circ$) during our observations we used the LOCI algorithm \citep{lafreniere2007b} to subtract the stellar PSF of our images. LOCI creates a reference PSF for each image from a linear combination of all other images observed at a different parallactic angles. The coefficients of the combination are optimized inside different subsections of the image independently so that the residual noise within each subsection is minimized. We refer the reader to the original paper for a more detailed description of LOCI. By scaling and inserting the PSF of the unsaturated images as fake planets with known brightness in the raw frames and retrieving them in the final image we did a small parameter study to optimize the LOCI parameters for our purposes. The best results in terms of planet contrast and S/N\footnote{A description how we define planet contrast and S/N is given in section~\ref{analysis}.} were achieved with the following LOCI parameters which we used for the final analyses: FWHM=4.5 px, $N_\delta$=0.75, $dr$=3, $N_A$=200.  We note that  we used LOCI on each individual image frame and not on stacked images as the former approach provided better detection performances. The data for each hemisphere was reduced separately. In a last step, we cut vertically through the center of each PSF-subtracted frame and saved only the high-contrast, i.e., left-hand, side of the PSF. For both hemispheres these images were then rotated to the same field orientation and averaged to create our final image. No additional filtering was applied to our data.

\section{Results \& Analysis}\label{analysis}
\begin{figure}
\centering
\epsscale{1}
\plottwo{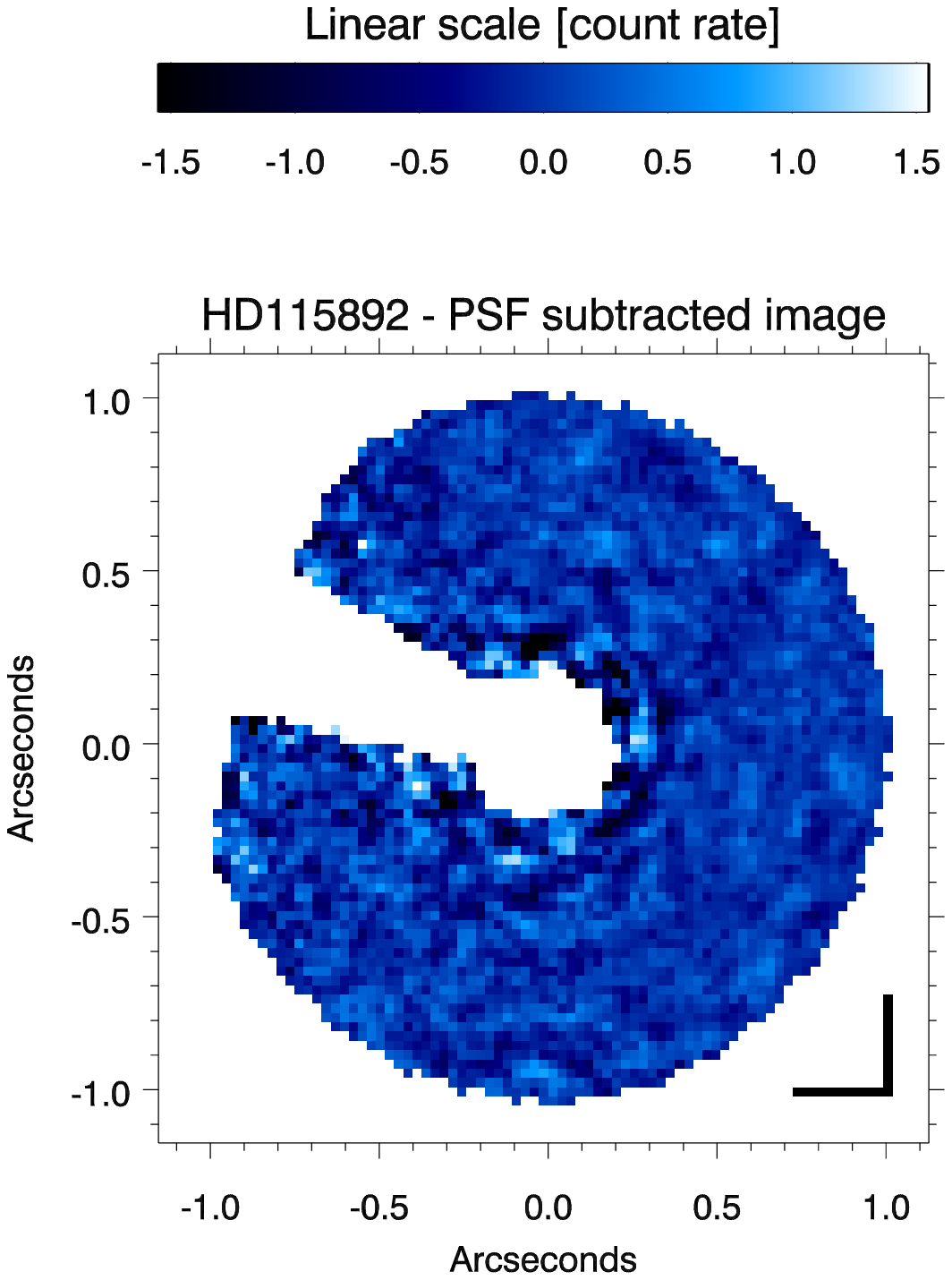}{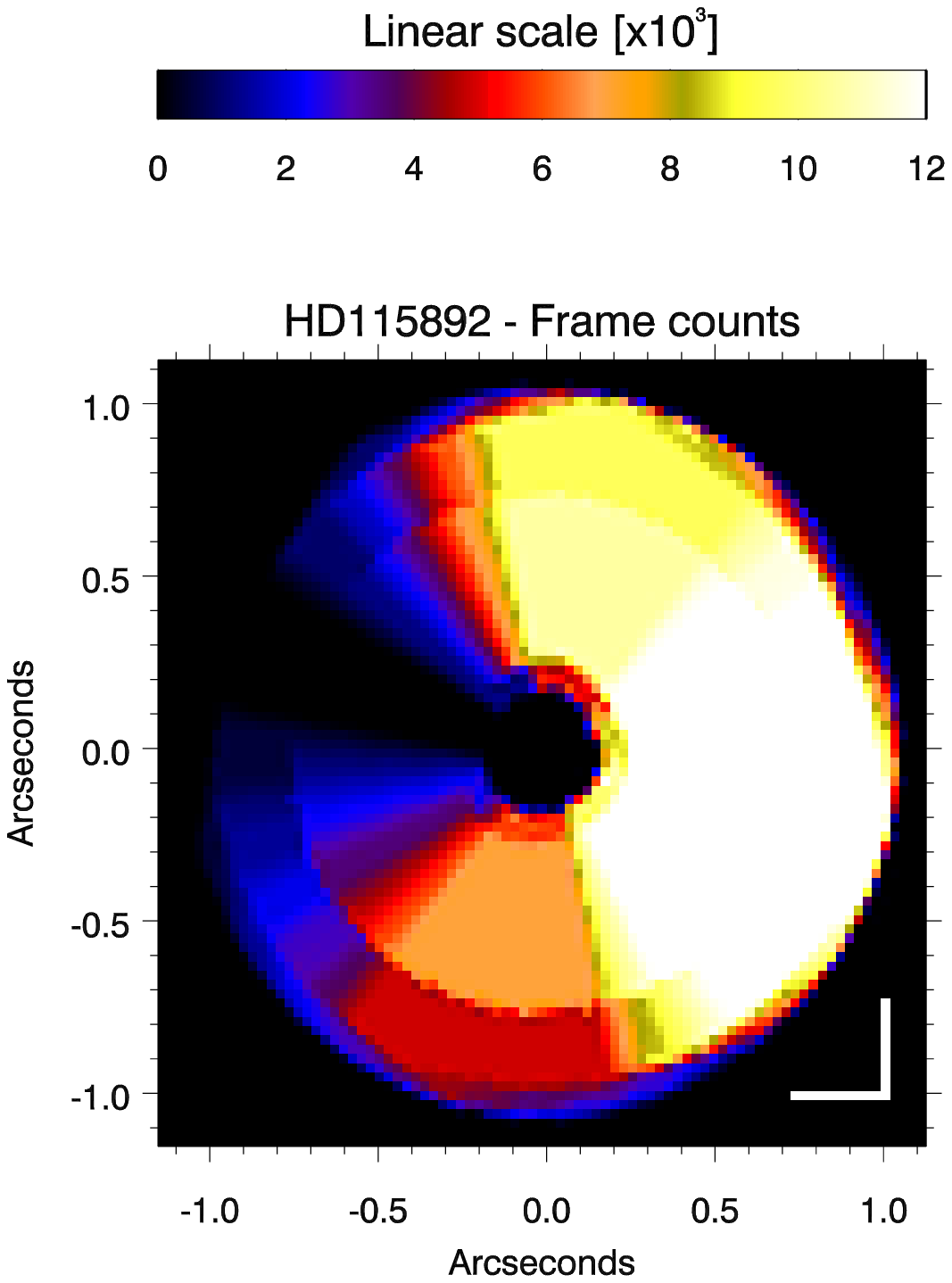}
\plottwo{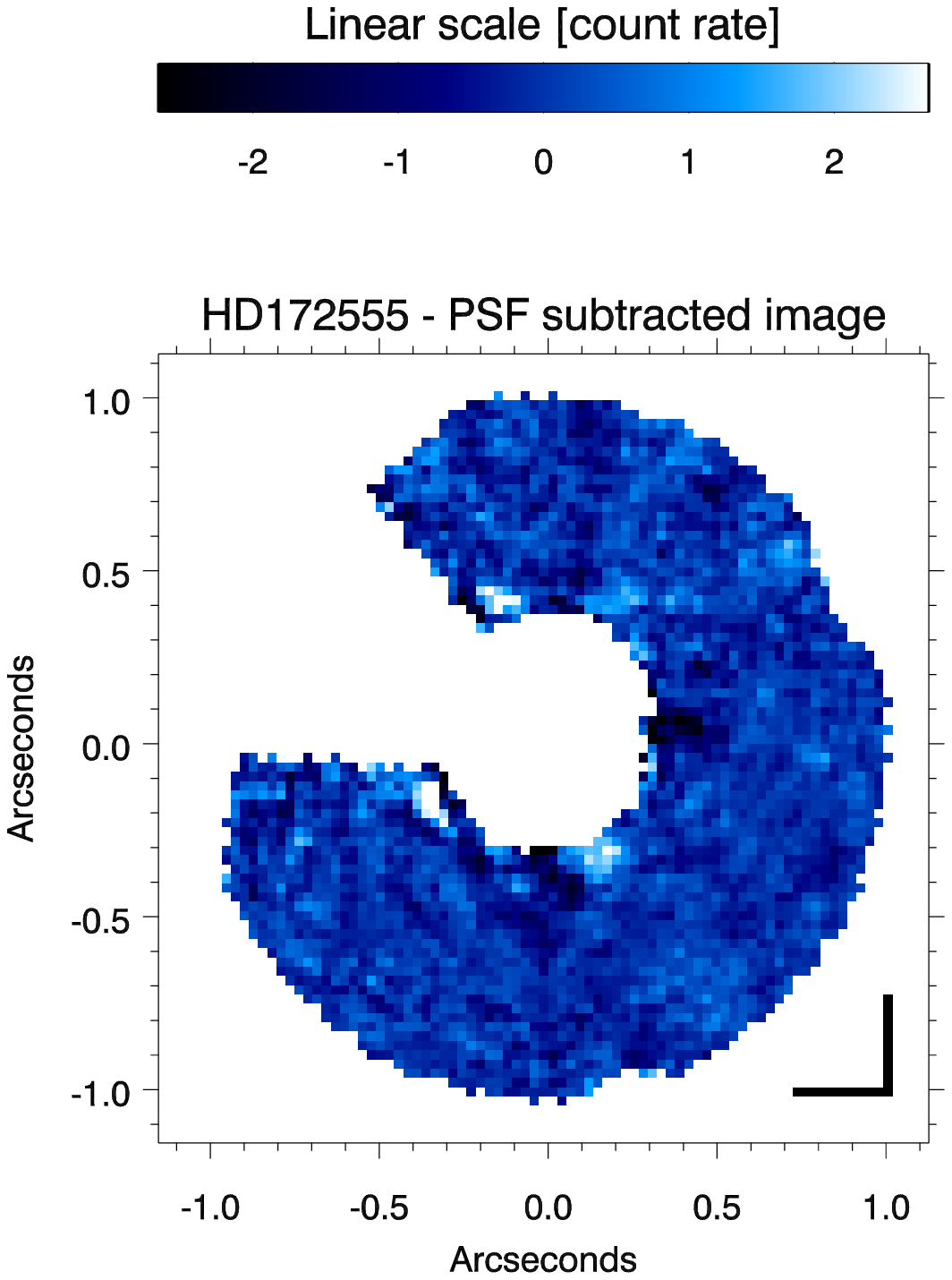}{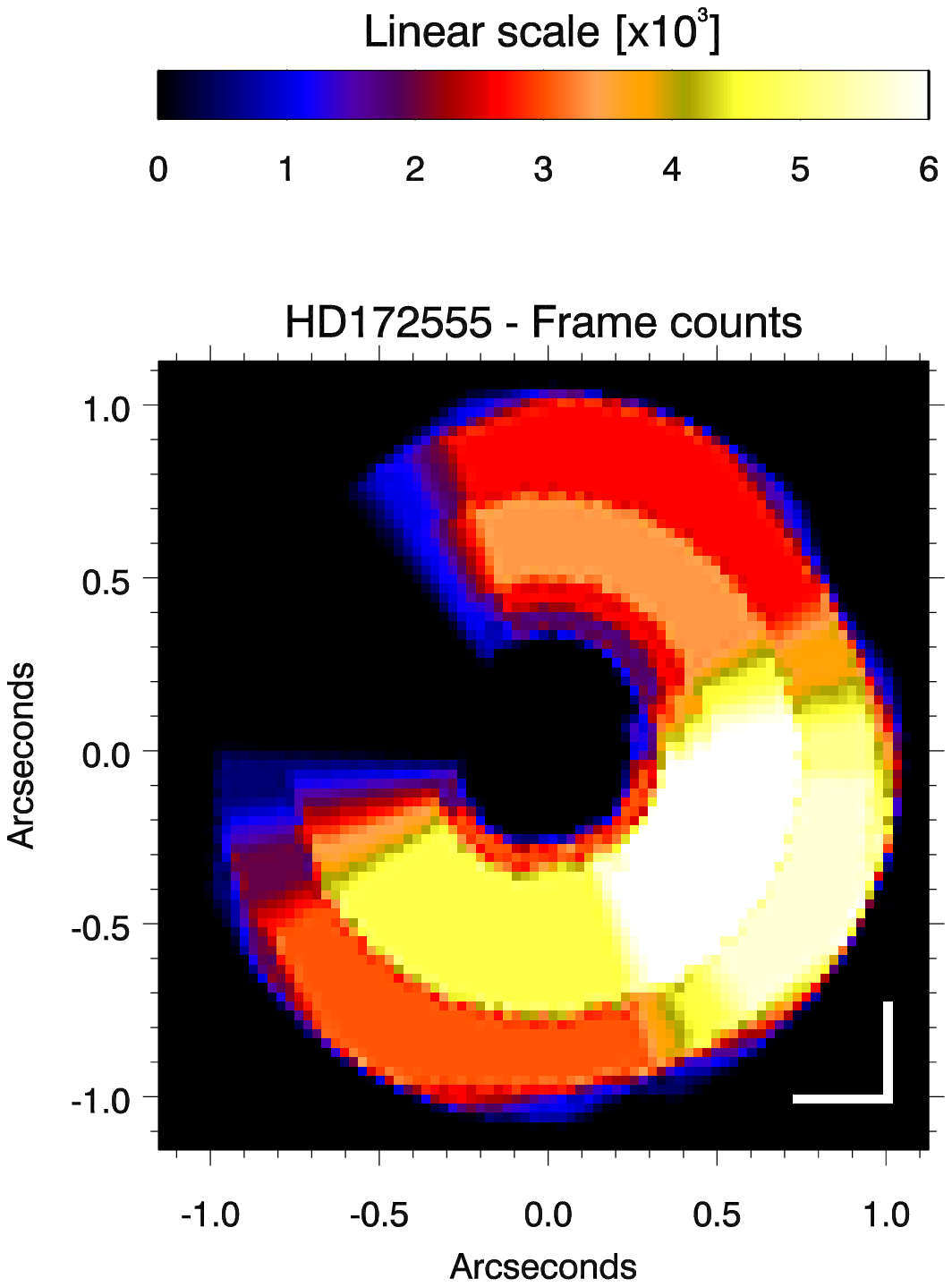}
\caption{Final PSF subtracted images of HD115892 and HD172555 (left column) and corresponding images showing the number of frames that were combined for a given position (right column). All images have a linear scale. The pixel units of the PSF subtracted images is count rate and the stretch ranges from -5$\sigma$ to +5$\sigma$ where $\sigma$ denotes the standard deviation of the counts in the background. We computed $\sigma$ in those regions in the image where more than the mean number of frames where combined (i.e., 7710 frames for HD115892 and 3494 frames for HD172555). Due to an error during the rotation of the camera we lack data for a wedge in the North-East quadrant around both targets (North is always up and East to the left).
\label{images}}
\end{figure}

In Figure~\ref{images} we show the final PSF-subtracted images for HD115892 and HD172555 in the left column. Due to the rotation error described above we lack data in a wedge in the North-East quadrant of both objects. On the opposite side of the central star, however, a wedge of the same size was covered during the observations of both hemispheres. The right column of Figure~\ref{images} shows the number of frames that were eventually combined for a given position for the final image. Our final images probe regions as close as $\sim$0.3$''$ (5.5 AU) and $\sim$0.4$''$ (11 AU) around HD115892 and HD172555, respectively, and out to 1$''$ as the maximum distance around each target. Since objects tend to slightly drift across the detector in NACO's pupil stabilized mode, we could only combine the maximum number of frames in the innermost $\sim$0.8$''$ (see, right column Figure~\ref{images}).


We did not detect any faint companions in our final images. However, by inserting and retrieving fake planets we can determine the sensitivity of our observations and put constraints on the maximum brightness of potential non-detected companions. Since LOCI can significantly reduce the flux of any detected point source, detection limits need to be based on the retrieval of fake companions and cannot be derived solely from the noise in the final image. A complicating factor in our case is the inhomogeneous sensitivity across the final images. We decided to put fake planets in regions where only data from one hemisphere is combined and not in the overlapping regions. This approach is representative for a "typical" APP observing run and the results are representative for typical APP detection limits. We used the data of hemisphere 1 for HD115892 and of hemisphere 2 for HD172555. For HD115892, 10681 and 9615 frames were combined for separations $\le$0.8$''$ and $>$0.8'', respectively. For HD172555 
we could combine 4697 and 3046 frames in these regions. As fake planets we used for each target the median-combined PSF of an unsaturated data set and scaled it to different contrast ratios based on the average count rate of the unsaturated images and the difference in exposure time between the unsaturated and the saturated images. These fake planets with known brightness were then inserted in the individual sky-subtracted raw frames at different radii taking into account the field rotation that occurred between the exposures. Finally, we repeated the data reduction described above and determined the S/N of fake planets that we recovered in the final image. We did aperture photometry on the recovered planets and compared it to the standard deviation of background pixels in an annulus centered on the central star. This annulus had the same radial distance as the planet and a width twice as wide as the aperture radius. We excluded those regions in the annulus where fewer frames were combined than at the position of the planet, and we excluded the region around the planet itself (i.e., 3 FWHM centered on the planet) as LOCI can create artificial 'holes' left and right of a detected point source. The S/N of the fake planet can then be expressed as 

\begin{equation}
S/N=F_{\rm pl}/(\sigma\cdot\sqrt{\pi r_{\rm ap}^2})
\end{equation}

 with $F_{\rm pl}$ being the flux of the planet, $\sigma$ the standard deviation of the pixels in the annulus (both measured in 'count rate') and $r_{\rm ap}$ the aperture radius. We inserted fake planets with a contrast between 9 and 11 mag in the HD115892 data and between 8 and 9 mag in the HD172555 data and computed the S/N for two aperture sizes (2 and 3 pixels radius). The final 5$\sigma$ contrast limit for a given separation was then derived by averaging the S/N in both apertures, taking those fake planets where the averaged S/N was the lowest but $\ge$5 and extrapolating the contrast of the planet to a value that would correspond to a 5$\sigma$ detection. We emphasize that we did not apply any sort of filtering or background smoothing to our data which makes our final S/N estimates rather conservative. Also the optimized extraction of a PSF template could lead to the robust detection of fainter companions.   

\begin{figure}
\centering
\epsscale{0.8}
\plottwo{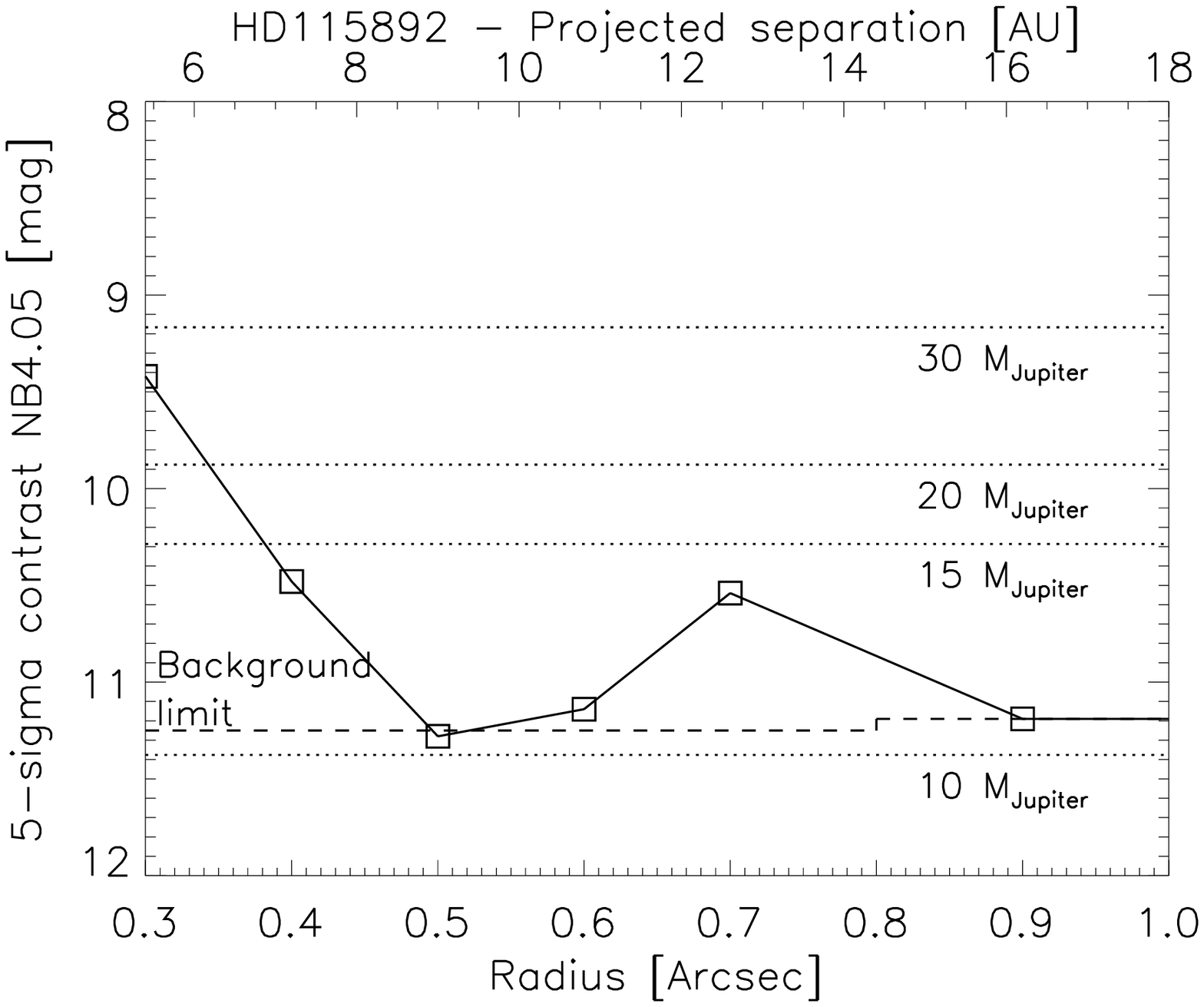}{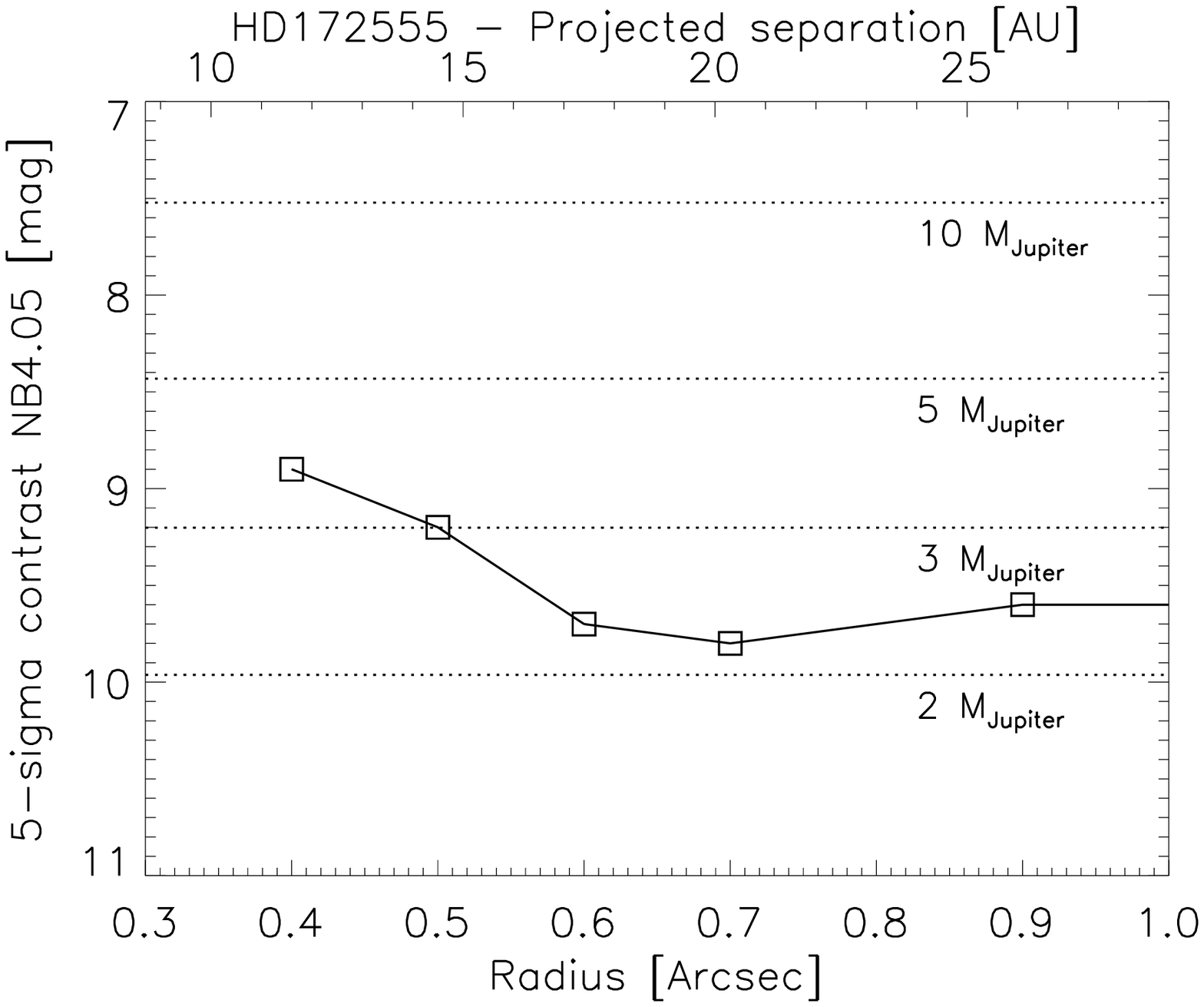}
\caption{5$\sigma$ detection limits (squares) given as magnitude contrast in the NB4.05 filter as a function of radial separation from the host star for HD115892 (left panel) and HD172555 (right panel). The lower x-axes show the radial separation in arcsec while the upper ones depict the projected separation in AU. Due to the larger field rotation in the dataset for HD115892 we can probe inner working angles as small as 0.3$''$. Overplotted are the expected contrast for planets with different masses (dotted lines). For HD115892 we also plot the measured background limit for our observations (dashed line). See text for more details. 
\label{detection_limits}}
\end{figure}

In Figure~\ref{detection_limits} we show the final 5$\sigma$ detection limits for both objects between 0.3--1.0$''$. Overplotted are detectable mass limits for a given contrast and the age of the star (Table~\ref{targets}). These mass limits are derived from the DUSTY and COND evolutionary models \citep{chabrier2000,baraffe2003}. We use the COND models for objects with effective temperatures below $\sim$1700 K and the DUSTY models for hotter objects. For the 350 Myr old object HD115892 our data reach a contrast between $\sim$10.5--11.3 mag at angular separations between 0.4--1.0$''$ (7--18 AU). This contrast corresponds to detectable mass limit between 10--15 M$_{\rm Jup}$. At 0.3$''$ ($\sim$5.5 AU) the contrast is $\sim$9.4 mag and we are still sensitive to objects with masses $\gtrsim$25 M$_{\rm Jup}$. For the 12 Myr HD172555 system the contrast is $\sim$9.2--9.8 mag at separations between 0.5--1.0$''$ (15--29 AU) which corresponds to mass limits of 2--3 M$_{\rm Jup}$. At 0.4$''$ ($\sim$11 AU) the achieved contrast is $\sim$8.9 mag and we are still sensitive to objects with $\gtrsim$4 M$_{\rm Jup}$. Due to the smaller field rotation for this object we can not probe IWA $\le$0.3$''$. 



Both our datasets have comparable total integration times and factoring in the apparent brightness of the stars both curves are comparable in terms of detectable brightness for potential companions. In addition, both contrast curves are relatively flat for separations $\gtrsim$0.5$''$. This suggests that the APP achieves close to background limited performance for these separations. We computed the expected background limit for the HD115892 data set based on the sky noise in individual frames far away from the star. The dashed line in the left panel in Figure~\ref{detection_limits} shows the result and confirms that our data are indeed (mostly) limited by the background and not by the contrast. Due to the lack of appropriate dark frames we could not repeat this exercise for HD172555. 


Given the non-homogeneous data coverage in azimuth the detection limits derived above vary between different positions around each object. 
To estimate the global detection limits we only consider those regions where we have combined at least half as many frames as for the analysis. For HD115892 we then have to exclude an azimuthal wedge between $\sim$22$^\circ$--134$^\circ$ (East of North), and for  HD172555 it's regions between $\sim$17$^\circ$--107$^\circ$ (see also right column Figure~\ref{images}). In all the other parts of the images the detection limits shown in Figure~\ref{detection_limits} apply with a significance of $\ge$3.5$\sigma$ with the lowest significance applying only in very small wedges directly adjacent to the excluded parts. 

\section{Discussion}
\citet{lagrange2009c} used radial velocity to search for planetary mass companions to both of our targets. Although they did not find any they could put some constraints on the occurrence of massive planets in short period orbits. For HD115892 they could exclude objects more massive than 1.7, 3.8 and 100.0 M$_{\rm Jup}$ in 3, 10, and 100 day orbits, respectively, with $>$99\% confidence. These orbital periods correspond to semi-major axes of $\sim$0.06, 0.13, and 0.62 AU assuming circular orbits. For HD172555 the same confidence level was achieved for objects more massive than 11.3 M$_{\rm Jup}$ in a 3 or 10 day orbit  (i.e., $\sim$0.05 AU and $\sim$0.11 AU), respectively. For a 100 day orbit (i.e., at 0.53 AU) the detection limit was 33 M$_{\rm Jup}$. In addition, \citet{biller2007} used NACO in spectral differential imaging (SDI) mode to search for low-mass companion around HD172555. While our data are more sensitive in the innermost 1$''$ (i.e., for separations $<$30 AU), their data cover regions out to 2$''$ and they could have detected objects with masses $\gtrsim$5 M$_{\rm Jup}$ for separation between $\sim$30--60 AU. 


In combination with the other studies our data put stringent constraints on the existence of giant exoplanets around HD115892 and HD172555 interior and exterior to the debris disks. Although our data probe regions very close to the assumed location of the debris disks (4--6 AU), our hypothesis that dynamical interactions between a planet and the debris disk could have led to a recent collision of planetesimal lacks direct observational support. We note, however, that planets (or planetary systems) with masses below our detection limits are certainly able to dynamically shape debris disks and influence their evolution \citep[see, e.g.,][]{raymond2011}.

Our data demonstrate that the APP opens up a new parameter space for direct imaging of exoplanets by pushing the background limit significantly closer to the star. A comparison to surveys carried out in the H band shows that \citet{chauvin2010} and \citet{lafreniere2007} typically reached a contrast of $\sim$10 mag and $\sim$9.5 mag at a separation of 0.5$''$, respectively. For HD115892 our contrast is $\gtrsim$11 mag at the same separation. A similar contrast has been reported by the NICI campaign at the Gemini observatory \citep{chun2008} also operating in the H band, but a more direct comparison of the contrast performance is limited due to different integration times and different target stars with different brightnesses. However, since planetary mass objects appear red in the infrared, NACO/APP has an advantage when it comes to the final detectable mass limits, because it works  in the L band and not in the H band. 

\section{Summary \& Conclusions}
We presented the first observations combining NACO's Apodizing Phase Plate coronagraph with Angular Differential Imaging to search for faint companions to the young debris disk host stars HD115892 and HD172555 in the NB4.05 filter. Our conclusions are as follows:
\begin{itemize}
\item We did not detect any point sources but achieved the following detection limits: 
For HD115892 we could have detected objects with a contrast of $\sim$10.5--11.3 mag (corresponding to 10--15 M$_{\rm Jup}$) at angular separations between 0.4--1.0$''$ (7--18 AU). At 0.3$''$ ($\sim$5.5 AU) the detection limit was a contrast of $\sim$9.4 mag ($\gtrsim$25 M$_{\rm Jup}$). For HD172555 we reached a contrast of $\sim$9.2--9.8 mag (2--3 M$_{\rm Jup}$) at separations between 0.5--1.0$''$ (15--29 AU). At 0.4$''$ ($\sim$11 AU) the detection limit was a contrast of 8.9 mag ($\sim$4 M$_{\rm Jup}$). These limits are $\ge$3.5$\sigma$ limits. We do not have data in an azimuthal wedge between $\sim$22$^\circ$--134$^\circ$  (East of North) for HD115892 and between $\sim$17$^\circ$--107$^\circ$ for HD172555.

\item While current/previous high-contrast imaging campaigns carried out in the H band are contrast limited at small IWA, our data are mostly background limited for separations $\gtrsim$0.5$''$ in the thermal infrared. 

\item Taking advantage of the red H--L color of planetary mass objects, NACO/APP is capable of detecting cooler planets (i.e., lower mass or older planets) compared to observations in the near-infrared for a given contrast. 
\end{itemize}
NACO/APP is currently a superior combination to search for planets at unprecedented small IWA in particular around bright targets. 
And even when the next generation high-contrast imaging instruments such as SPHERE and GPI come online, with its unique L-band capabilities NACO/APP can help to characterize at least a certain subset of the exoplanets these instrument will find in the near-infrared.



\acknowledgments
This research has made use of the SIMBAD database, operated at CDS, Strasbourg, France. We thank D. Lafreni\`ere for kindly allowing us to adapt his LOCI source code. We thank  C. Thalmann for his support setting up the data reduction pipeline.  M. Janson and I. Baraffe kindly provided us with the evolutionary models in the NB4.05 filter. We are indebted to U. Wehmeier and the ESO staff on Paranal, in particular J. O'Neal, for their support during the observations. 



{\it Facilities:} \facility{VLT:Yepun (NACO)}

\clearpage

\clearpage

\begin{deluxetable}{lll}
\tablecaption{Basic properties of target stars based on \citet{su2006} and \citet{wyatt2007} and references therein.
\label{targets}}           
\tablewidth{0pt}
\tablehead{
\colhead{Parameter}  & \colhead{HD115892}  & \colhead{HD172555} 
}
\startdata
RA (J2000) & $13^h20^m35^s.82$  & $18^h45^m26^s.90$ \\
DEC (J2000) & $-36^\circ42'44''.26$ & $-64^\circ52'16''.53$ \\
Apparent mag. in L  &	2.68 mag\tablenotemark{a} & 4.28 mag\tablenotemark{b}\\
Spectral type & A2V & A5IV/V\tablenotemark{c} \\
Mass & 2.5 M$_\sun$ & 2.0 M$_\sun$\\
Age & 350 Myr&  12 Myr\\
Distance & 18 pc & 29 pc \\
Debris disk radius & 6 AU & 4 AU \\ 
\enddata
\tablenotetext{a}{\citet{morel1978}}
\tablenotetext{b}{Based on K-band magnitude from \citet{cutri2003} and a K-L color of 0.02 mag for an A5V star (see: http://www.jach.hawaii.edu/UKIRT/astronomy/utils/temp.html).}
\tablenotetext{c}{\citet{gray2006} found a spectral type of A7V.}
\end{deluxetable}

\begin{deluxetable}{lllll}
\tablecaption{Summary of deep imaging observations in pupil tracking mode.
\label{observations}}           
\tablewidth{0pt}
\tablehead{
\colhead{Parameter}  & \colhead{HD115892} & \colhead{HD115892} & \colhead{HD172555} & \colhead{HD172555} \\
\colhead{} & \colhead{1st hemisphere} & \colhead{2nd hemisphere} & \colhead{1st hemisphere} & \colhead{2nd hemisphere} \\
}
\startdata
UT start & 02h:48m:19.28s & 05h:03m:52.30s & 06h:44m:42.80s & 08h:26m:32.36s \\
UT end &  04h:43m:13.37s & 06h:16m:41.86s  & 08h:05m:19.54s & 10h:11m:23.80s \\
NDIT $\times$ DIT\tablenotemark{a}  & 200 $\times$ 0.5 s & 200 $\times$ 0.5 s  & 200 $\times$ 1.2 s & 200 $\times$ 1.2 s\\
NINT\tablenotemark{b} & 54 & 36 & 18 & 24 \\
Parallactic angle start & -78.99$^{\circ}$ & -10.04$^{\circ}$  & -71.76$^{\circ}$ & -43.78$^{\circ}$\\
Parallactic angle end & -28.52$^{\circ}$ & 53.63$^{\circ}$ & -48.88$^{\circ}$ & -7.81$^{\circ}$\\
Airmass & 1.19\dots1.03 & 1.02\dots1.05 & 1.71\dots1.46 & 1.42\dots1.31 \\
Typical DIMM seeing [$\lambda$=500 nm]&  0.6$''$\dots0.8$''$ & 0.5$''$\dots0.7$''$ & 0.5$''$\dots0.6$''$ & 0.5$''$\dots0.9$''$ \\
$\langle{EC}\rangle_{\rm mean}$ / $\langle{EC}\rangle_{\rm min}$ / $\langle{EC}\rangle_{\rm max}$\tablenotemark{c} & 46.2 / 8.5 / 58.6 \% & 49.0 / 14.6 / 64.1 \% & 36.97 / 23.3 / 48.8 \% & 47.1 / 25.2 / 58.9 \% \\ 
$\langle \tau_0 \rangle_{\rm mean}$ / $\langle \tau_0 \rangle_{\rm min}$ / $\langle \tau_0 \rangle_{\rm max}$\tablenotemark{d} & 9.0 / 5.2 / 13.1 ms & 9.3 / 5.6 / 13.0 ms & 4.1 / 2.9 / 5.3 ms & 6.0 / 3.7 / 8.5 ms \\
PA$_{\rm camera}$\tablenotemark{e} & -170.00$^{\circ}$\dots-121.00$^{\circ}$& 78.95$^{\circ}$\dots141.73$^{\circ}$ & -162.77$^{\circ}$\dots-141.06$^{\circ}$& 45.21$^{\circ}$\dots79.76$^{\circ}$\\
\enddata
\tablenotetext{a}{NDIT = Number of detector integration times (i.e., number of individual frames); DIT = Detector integration time (i.e., single frame exposure time).}
\tablenotetext{b}{NINT = Number of data cubes.}
\tablenotetext{c}{Average, minimum and maximum value of the coherent energy of the PSF in data cubes. Calculated by the Real Time Computer of the AO system.}
\tablenotetext{d}{Average, minimum and maximum value of the coherence time of the atmosphere in data cube. Calculated by the Real Time Computer of the AO system.}
\tablenotetext{e}{Position angle of camera adaptor at the beginning of exposure.}
\end{deluxetable}

\end{document}